\documentclass[11pt,amsmath,amssymb,]{revtex4-2}

\usepackage{graphicx}
\usepackage{dcolumn}
\usepackage{bm}
\usepackage[colorlinks=true]{hyperref}

\begin{document}


\title{Kapitza-resistance-like exciton dynamics in atomically flat MoSe$_{2}$-WSe$_{2}$ lateral heterojunction}

\author{Hassan Lamsaadi$^{1}$}
\author{Dorian Beret$^{2}$}
\author{Ioannis Paradisanos$^{2,9}$}
\author{Pierre Renucci$^2$}
\author{Delphine Lagarde$^2$}
\author{Xavier Marie$^2$}
\author{Bernhard Urbaszek$^{2,5}$}
\author{Ziyang Gan $^{3}$}
\author{Antony George$^{4}$}
\author{Kenji Watanabe$^7$}
\author{Takashi Taniguchi$^8$}
\author{Andrey Turchanin$^{3,4}$}
\author{Laurent Lombez$^2$}
\email{laurent.lombez@insa-toulouse.fr}
\author{Nicolas Combe$^1$}
\author{Vincent Paillardi$^1$}
\author{Jean-Marie Poumirol$^1$}
\email{jean-marie.poumirol@cemes.fr}

\affiliation{\small$^1$CEMES-CNRS, Universit\'e de Toulouse, Toulouse, France}
\affiliation{\small$^2$Universit\'e de Toulouse, INSA-CNRS-UPS, LPCNO, 135 Avenue Rangueil, 31077 Toulouse, France}
\affiliation{\small$^3$Friedrich Schiller University Jena, Institute of Physical Chemistry, 07743 Jena, Germany}
\affiliation{\small$^4$Abbe Centre of Photonics, 07745 Jena, Germany}
\affiliation{\small$^5$Ulm University, Central Facility of Electron Microscopy, D-89081 Ulm, Germany}
\affiliation{\small$^5$Institute of Condensed Matter Physics, Technische Universität Darmstadt, Darmstadt, Germany}
\affiliation{\small$^7$Research Center for Functional Materials, National Institute for Materials Science, 1-1 Namiki, Tsukuba 305-0044, Japan}
\affiliation{\small$^8$International Center for Materials Nanoarchitectonics, National Institute for Materials Science, 1-1 Namiki, Tsukuba 305-0044, Japan}
\affiliation{\small$^9$Institute of Electronic Structure and Laser, Foundation for Research and Technology-Hellas, Heraklion, 70013, Greece}

\begin{abstract}
Being able to control the neutral excitonic flux is a mandatory step for the development of future room-temperature two-dimensional excitonic devices. Semiconducting Monolayer Transition Metal Dichalcogenides (TMD-ML) with extremely robust and mobile excitons are highly attractive in this regard.  However,  generating an efficient and controlled exciton transport over long distances is a very challenging task. Here we demonstrate that an atomically sharp TMD-ML lateral heterostructure (MoSe$_{2}$-WSe$_{2}$) transforms the isotropic exciton diffusion into a unidirectional excitonic flow through the junction. Using tip-enhanced photoluminescence spectroscopy (TEPL) and a modified exciton transfer model, we show a discontinuity of the exciton density distribution on each side of the interface. We introduce the concept of exciton Kapitza resistance, by analogy with the interfacial thermal resistance referred to as Kapitza resistance. By comparing different heterostructures with or without top hexagonal boron nitride (hBN) layer, we deduce that the transport properties can be controlled, over distances far greater than the junction width, by the exciton density through near-field engineering and/or laser power density. This work provides a new approach for controlling the neutral exciton flow, which is key toward the conception of excitonic devices.
 \end{abstract}
\maketitle

\section{\label{sec:level1}Introduction}

Electronics relies on the control of the motion of charge carriers to process information. The losses caused by the charged particle current and the resulting need to improve the power efficiency have fueled lots of interest in recent years \cite{butov_excitonic_2017, 867687}. As it is based on the control of electrically neutral quasi-particles (excitons), insensitive to long-range Coulomb scattering mechanisms, excitronics is by nature much more power efficient as it presents only negligible ohmic losses \cite{high_control_2008, peng_long-range_2022}. Nevertheless, developing excitronic devices is challenging, as it requires the ability to control the neutral exciton properties, such as the recombination rates, diffusion length $(L_{D})$ and propagation direction, in an optically active medium at room temperature, without the help of external electric or magnetic fields \cite{high_exciton_2007, high_control_2008}. 

Owing to their promising optical properties, Transition Metal Dichalcognenide monolayers emerged as a highly versatile platform for excitonics system at the nanoscale \cite{wang_colloquium_2018, wang_electroluminescent_2018, novoselov_2d_2016, cheiwchanchamnangij_quasiparticle_2012}. In particular, their large exciton binding energy allows operating at room temperature. Due to their unique band properties, excitonic transport in TMD-MLs has led to the discovery of new fundamental phenomena such as a valley hall effect \cite{mak_valley_2014, ubrig_microscopic_2017, zhou_spin-orbit_2019}, or the observation of nonlinear behavior such as a halo in the spatial profile \cite{kulig_exciton_2018} or negative effective diffusion \cite{rosati_negative_2020}. With the increasing maturity of the field, several basic components necessary to control exciton information have been developed, like room temperature excitonic transistor in a Van Der Waals vertical heterostructure \cite{unuchek_room-temperature_2018}, or an excitonic diode able to filter excitons \cite{beret_exciton_2022, shanks_interlayer_2022}. However, the exciton diffusion is up to now mainly driven either by strain in TMD-ML \cite{dirnberger_quasi-1d_2021} or dielectric gradient engineering in TMD based vertical heterostructure \cite{shanks_interlayer_2022}.  Both approaches require complex architectures, with nanometric precision of the strain or the electrostatic potential over large distances (micrometers). The fabrication of those excitonic guides as well as their coupling with other circuit elements is thus very challenging. 
 
In this paper, we investigate experimentally and theoretically the effect of an atomically sharp MoSe$_{2}$-WSe$_{2}$ lateral heterostructure (LH) on exciton diffusion and distribution. We performed tip-enhanced photoluminescence (TEPL) spectroscopy  experiments, allowing sub-wavelength spatial resolution down to 30 nm, and developed an exciton transfer model. We show that the difference in the energy gap at the LH generates a discontinuity in the exciton density distribution. In steady state conditions, the presence of this discontinuity results in unique non-reciprocal exciton transport properties, taking place over distances far greater (two orders of magnitude) than the junction width and experimentally evidenced by: a highly asymmetric photoluminescence (PL) profile, the quenching of the WSe$_2$-related PL and an enhancement of the MoSe$_2$-related PL. Furthermore, by comparing the diffusion properties of fully hBN-encapsulated LH and hBN supported LH (without top hBN layer), we demonstrate that the diffusion properties of the LH can be tuned, by the generated exciton population either by increasing the laser power density or by modifying the optical near-field configuration (at constant laser power).

\title{\textbf{Experimental results  --} }
\begin{figure}
	\includegraphics*[width=8.5cm]{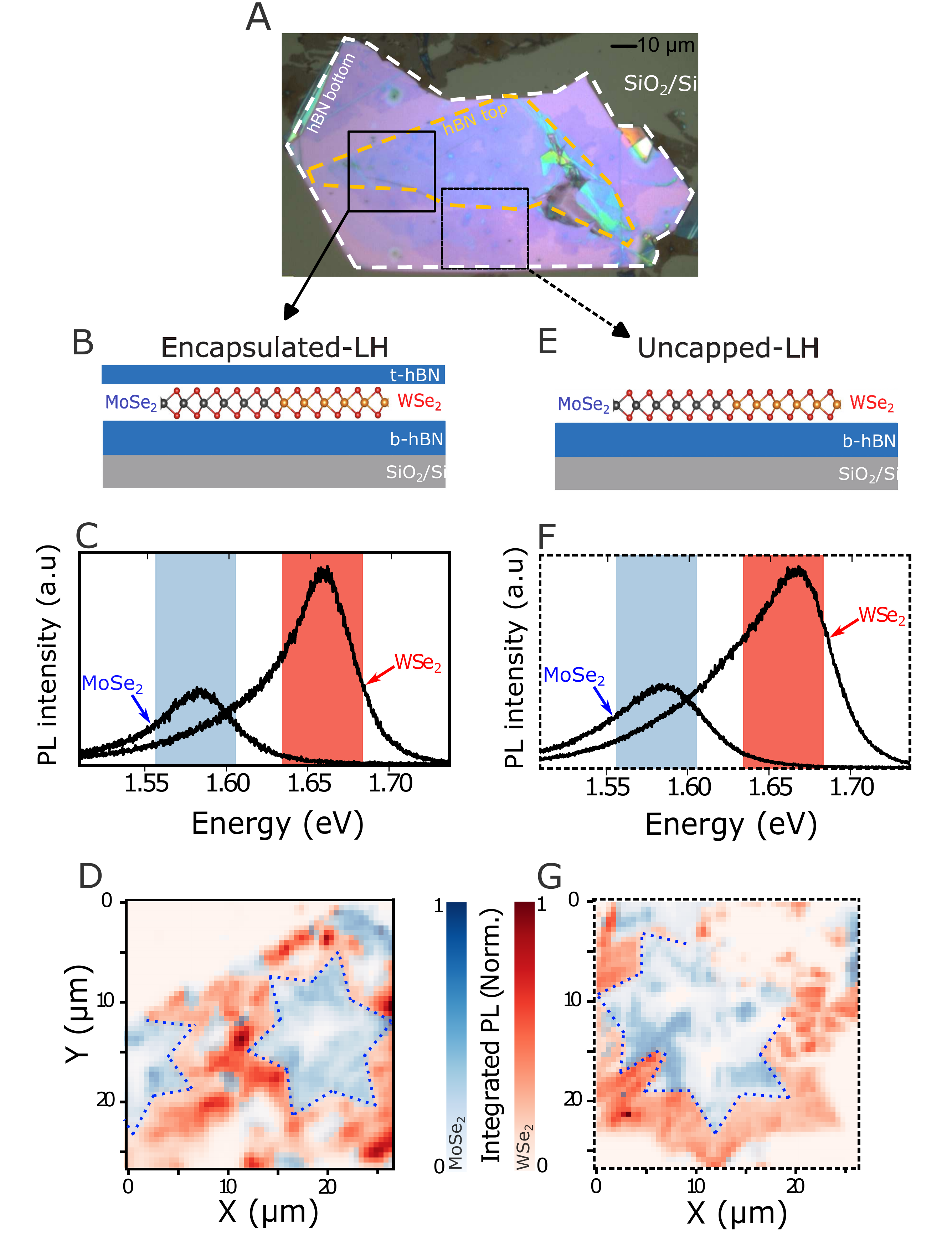} 
	\caption{ \textbf{A}. Optical image of the sample, the white (yellow) dashed contour shows the bottom (top) hBN flake boundaries. The continuous (respectively dashed) black square contour highlights the MoSe$_{2}$-WSe$_{2}$ LH, encapsulated in hBN (e-LH) region (respectively the hBN supported LH (un-LH) region). \textbf{B}. (\textbf{E}.) Schematic representation of the e-LH (un-LH).  \textbf{C}.  (\textbf{F}.) typical far-field PL spectra measured in the WSe$_{2}$ and MoSe$_{2}$ regions of e-LH  (un-LH). \textbf{D}. (\textbf{G}.) Spectrally integrated PL intensity maps of e-LH (un-LH). The PL intensity is obtained by integrating MoSe$_{2}$ (WSe$_{2}$) PL spectra over the spectra range represented by the blue (red) shaded area. PL spectra are recorded every 500 nm (step size), using a 633 nm excitation laser, 400 $\mu$W laser power. The dotted blue lines in \textbf{D}. and \textbf{G}. highlight the boundaries between the two materials.}
 \label{fig1}
\end{figure}

\section{\label{sec:level1}Experimental Results}

\subsection{\label{sec:level2}Sample preparation and characterization}
The high quality monolayer MoSe$_{2}$-WSe$_{2}$ LH is grown using a modified CVD method described in \cite{najafidehaghani_1d_2021}. We then use water-assisted deterministic transfer to pick up the LH from the as-grown substrate and transfer it on a supporting flake of exfoliated hBN on SiO$_{2}$/Si substrate. Finally, a second exfoliated hBN flake is deposited to cover the structure partially \cite{paradisanos_controlling_2020}. As a result, we obtain two distinct areas, as shown in Fig.\ref{fig1} \textbf{A}, a fully encapsulated area hBN/WSe$_{2}$-MoSe$_{2}$/hBN/SiO$_{2}$/Si (e-LH), and an uncapped area WSe$_{2}$-MoSe$_{2}$/hBN/SiO$_{2}$/Si (un-LH).  The dashed blue and yellow line in Fig. \ref{fig1} \textbf{A} highlight the boundaries of the bottom and top hBN flakes, respectively. 

Fig.\ref{fig1} \textbf{C} and \textbf{F} display the room temperature $\mu$-PL spectra measured on WSe$_{2}$ and MoSe$_{2}$, far from the junction on the encapsulated (e-LH) and uncapped (un-LH) regions, respectively. For the e-LH zone, the PL spectrum measured on MoSe$_2$ ML exhibits a pronounced peak centered around 1.575~eV with a full width at half maximum (FWHM) of 50 meV that, in agreement with previous room temperature measurements \cite{poumirol_unveiling_2020}, can be assigned to the neutral exciton (A$_{1s}^{MoSe_2}$). In the case of WSe$_2$, the PL spectrum is asymmetric, due to an intense peak linked to the A$_{1s}^{WSe_2}$ transition located around 1.65eV (FWHM $\sim$ 30 meV) and a second weaker peak $\sim$ 40 meV below A$_{1s}$, attributed to the spin-forbidden dark exciton (X$_D$) \cite{poumirol_unveiling_2020}. For the un-LH region, all previously described PL features appear nearly identical, with an increased intensity. Furthermore, we observe a slight increase of the broadening of all transitions, of the order of 5$\%$. One can also notice that all PL spectra are more asymmetric. This could be attributed to an increased contribution of the charged excitons (trions), as the unprotected sample can get chemically charged when exposed to ambient air. In any case, one can notice that the integrated PL intensity I$_{PL}^\infty$($A_{1s}^{WSe_2}$) measured far from the interface is $\approx$3 times stronger than  I$_{PL}^\infty$($A_{1s}^{MoSe_2}$). Fig.\ref{fig1} \textbf{D} and \textbf{G} are spectrally integrated $\mu$-PL intensity color maps. The red (respectively blue) color intensity is the value of integrated PL (shaded areas in \textbf{C} and \textbf{F}) from 1.620 eV to 1.650 eV (respectively from 1.550 eV to 1.580 eV). Fig.\ref{fig1} \textbf{D} (\textbf{G}) corresponds to the mapping of the encapsulated (uncapped) region. One can clearly see that the MoSe$_2$ layer is organized in 6 folded stars surrounded by WSe$_2$. The white regions, signature of low emission regions, have different origin. They correspond to bilayer inclusions when located at the center of MoSe$_2$ stars, but also reveal cracks or/and inclusions that are related to the transfer process. We used the far-field PL color maps in combination with Raman mappings to select interfaces away from any visible defects.

\subsection{\label{sec:level2} Near-field studies of the lateral heterojunction}

\begin{figure*}
	\includegraphics*[width=15cm]{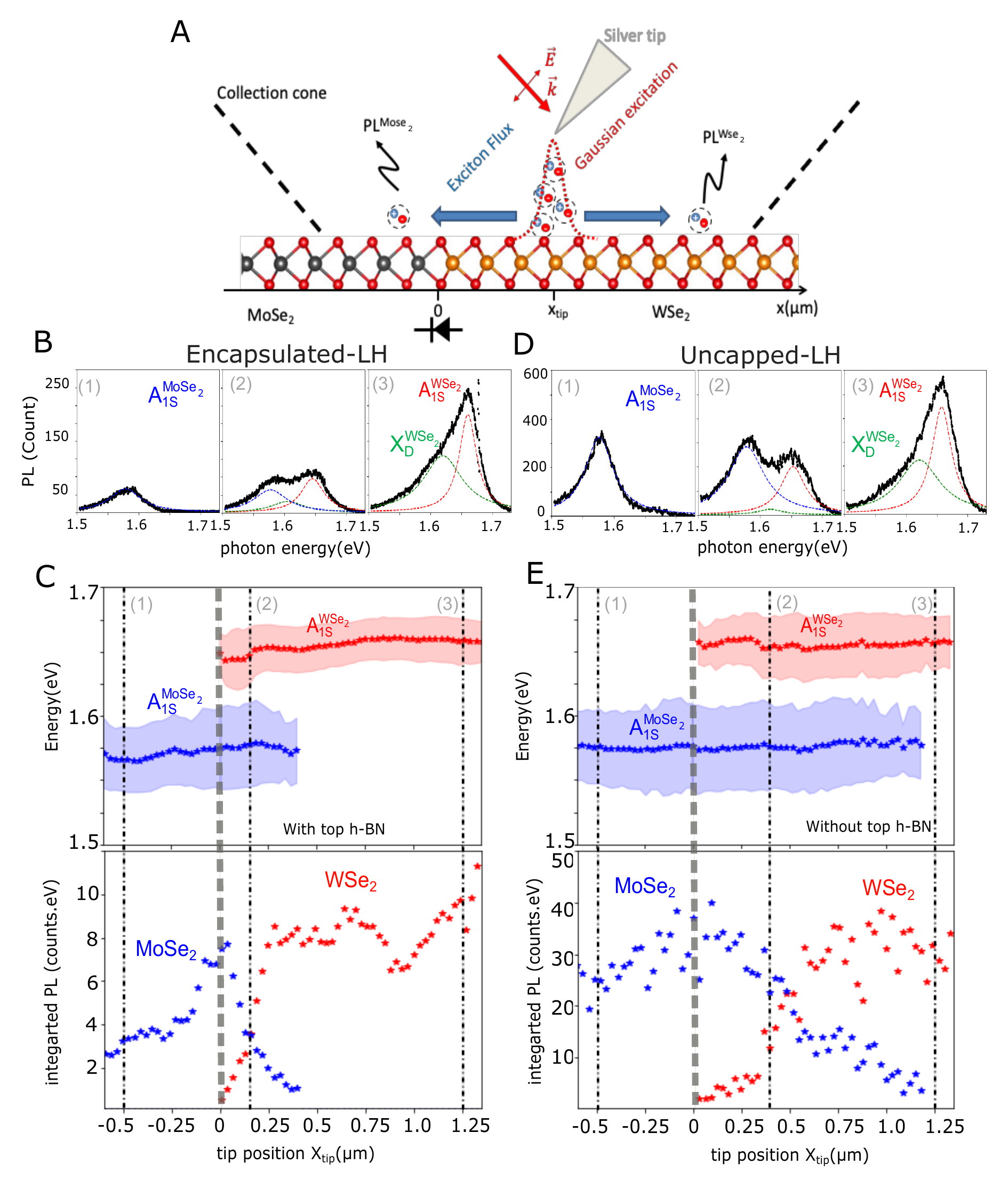} 
	\caption{ \textbf{A}. Schematic of the lateral heterojonction, TEPL measurement and the resulting excitonic diffusion properties. \textbf{B}. (\textbf{D}.) Typical TEPL spectra taken across the interface in e-LH (respectively un-LH) (1) 500 nm to the left of the interface, (2) 100 nm (300 nm) and (3) 1,25$\mu$m to the right of the interface. The excitonic contributions are fitted using individual Lorentzian function, neutral WSe$_2$ exciton in red (A$_{WSe_{2}}^{1s}$), neutral MoSe$_2$ exciton in blue (A$_{MoSe_{2}}^{1s}$) and the dark exciton (out-of-plan) in green (X$_{WSe_{2}}^{D}$). \textbf{C}. (\textbf{E}.) Top: Energy and FWHM of each Lorentzian peak obtained from the fitting procedure as shown in \textbf{B}. (\textbf{D}). Bottom: Amplitude of each Lorentzian peak obtained from the fitting procedure as shown in \textbf{B}. (\textbf{D}.) The red and blue stars indicate A$_{WSe_{2}}^{1s}$ and A$_{MoSe_{2}}^{1s}$, respectively.
	} 
	\label{fig2}
\end{figure*}

In this type of LH, the junction between the two materials WSe$_2$ and MoSe$_2$ is extremely sharp, down to $\sim 3$ nm as measured by electron microscopy \cite{beret_exciton_2022}. Therefore, in order to gain insight into the transport properties around the junction, we use a sub-wavelength resolution tool,  TEPL imaging and spectroscopy. Fig.\ref{fig2} \textbf{A.} displays a schematic of the experimental set-up, showing the linearly polarized laser excitation of  633 nm wavelength focused  onto the apex of an atomic force microscope (AFM) silver-coated tip. The exciton generation profile, is then linked to the electric field exaltation under the tip, a Gaussian profile with a full width at half maximum of $\sim 30$ nm, and a tip position (x$_{tip}$) controlled with AFM resolution. As illustrated in the Fig. \ref{fig2} \textbf{A.} excitons diffuse away from the excitation spot over long distances before recombining. The collection of emitted photons is ensured by a long working distance high numerical aperture (0.7 NA) microscope objective, with a fully open collection aperture. The position of the junction is determined very precisely ($\sim$ 30 nm) using tip enhanced Raman spectroscopy (TERS), as described in \cite{beret_exciton_2022}. 

TEPL line scans with 30 nm step size are obtained by scanning the tip along a line perpendicularly to the LH junction. Fig.\ref{fig2} \textbf{B} and \textbf{D} display typical TEPL spectra measured for three tip positions ($x_{tip}$) along the measured line for both encapsulated and uncapped samples. The spectra (1) and (3) recorded far from the interface (500 nm inside the MoSe$_2$ region and 1.25 $\mu$m inside the WSe$_2$ region) are  similar to the ones described in Fig.\ref{fig1}. The spectra of WSe$_{2}$ in locations (2) are recorded at 100 nm (resp. 300 nm) from the interface for the e-LH (resp. un-LH) sample. They show contributions of both MoSe2 and WSe2.

All TEPL spectra can be fitted using three Lorentzian peaks to account for the previously described relative contributions of A$_{1s}^{WSe_{2}}$ (dashed red line), X$_{WSe_{2}}^{D}$ (dashed green line) and  A$_{1s}^{MoSe_{2}}$ (dashed blue line). Fig.\ref{fig2} \textbf{C} and \textbf{E} display the result of this fitting procedure. The peak positions and FWHMs (shaded area) of the bright excitons A$_{1s}^{MoSe_{2}}$ and A$_{1s}^{WSe_{2}}$ are shown as a function of the tip position $x_{tip}$ in the upper panels, while the integrated PL intensity (I$_{PL}$) appears in the lower panels. For clarity, the contribution of  X$_{WSe_{2}}^{D}$ is not displayed (see supplement). First, we point out that the A$_{1s}^{MoSe_{2}}$ signature (blue star) is clearly observed while the excitation is taking place inside WSe$_2$, ($x_{tip}~>~0$, right side of the solid gray line in Fig.\ref{fig2} \textbf{C} and \textbf{E}). A contrario, no signature of A$_{1s}^{WSe_{2}}$ is observed when the excitation takes place in MoSe$_{2}$ ($x_{tip}~<~0$, left side of the solid gray line). This reveals that nonreciprocal filtering is taking place at the interface, allowing the excitons to cross the junction from WSe$_{2}$ to MoSe$_{2}$. The other transport direction being forbidden by the junction, in agreement with \cite{beret_exciton_2022, shimasaki_directional_2022}. The difference in the local dielectric environment between e-LH and un-LH has no impact on the diode-like effect. There is no visible influence on both the energy and FWHM of the PL spectra after crossing or being blocked at the junction. This suggests that, there is no alteration of the nature of the excitons in each TMD-ML near the vicinity of the junction.

It is very interesting to notice that, in both e-LH and un-LH, the TEPL intensity originating from WSe$_2$ (red stars) strongly decreases as the excitation takes place closer to the interface ($x_{tip} \rightarrow 0$). As a consequence, the resulting signal measured near the LH interface is strongly dominated by A$_{1s}^{MoSe_{2}}$ (see Fig.\ref{fig2} \textbf{C} and \textbf{E} bottom panels). In other words, even if the tip is located inside WSe$_2$, most of the generated excitons migrate through the LH junction into MoSe$_2$ before recombining. This behavior can only be explained by a strongly anisotropic transport, with the diffusion toward the junction becoming more efficient than the other directions. The second effect resulting from such an efficient exciton transfer A$_{1s}^{WSe_{2}}$ $\rightarrow$  A$_{1s}^{MoSe_{2}}$ is that as the tip approaches the interface from the WSe$_2$ side the MoSe$_2$ PL intensity I$_{PL}$(A$_{1s}^{MoSe_{2}}$) increases and nearly reaches the values of the integrated PL intensity of WSe$_2$ far from the interface (I$_{PL}^\infty$(A$_{1s}^{WSe_{2}}$)). Far from the junction, MoSe$_2$ is three times less bright than WSe$_2$, meaning that the junction is enhancing the MoSe$_2$ flake brightness at its vicinity.

Finally, a pronounced difference can be seen between the two systems, with A$_{MoSe_{2}}^{1s}$  signature extending $\sim 1.2~\mu$m away from the interface in un-LH versus $\sim$~400 nm for e-LH. Both distances being considerably larger than the physical junction width ($\approx$ 3 nm \cite{beret_exciton_2022})
\vspace{0.3cm} 
\begin{figure}[!t]
\centering
	\includegraphics[width=8cm]{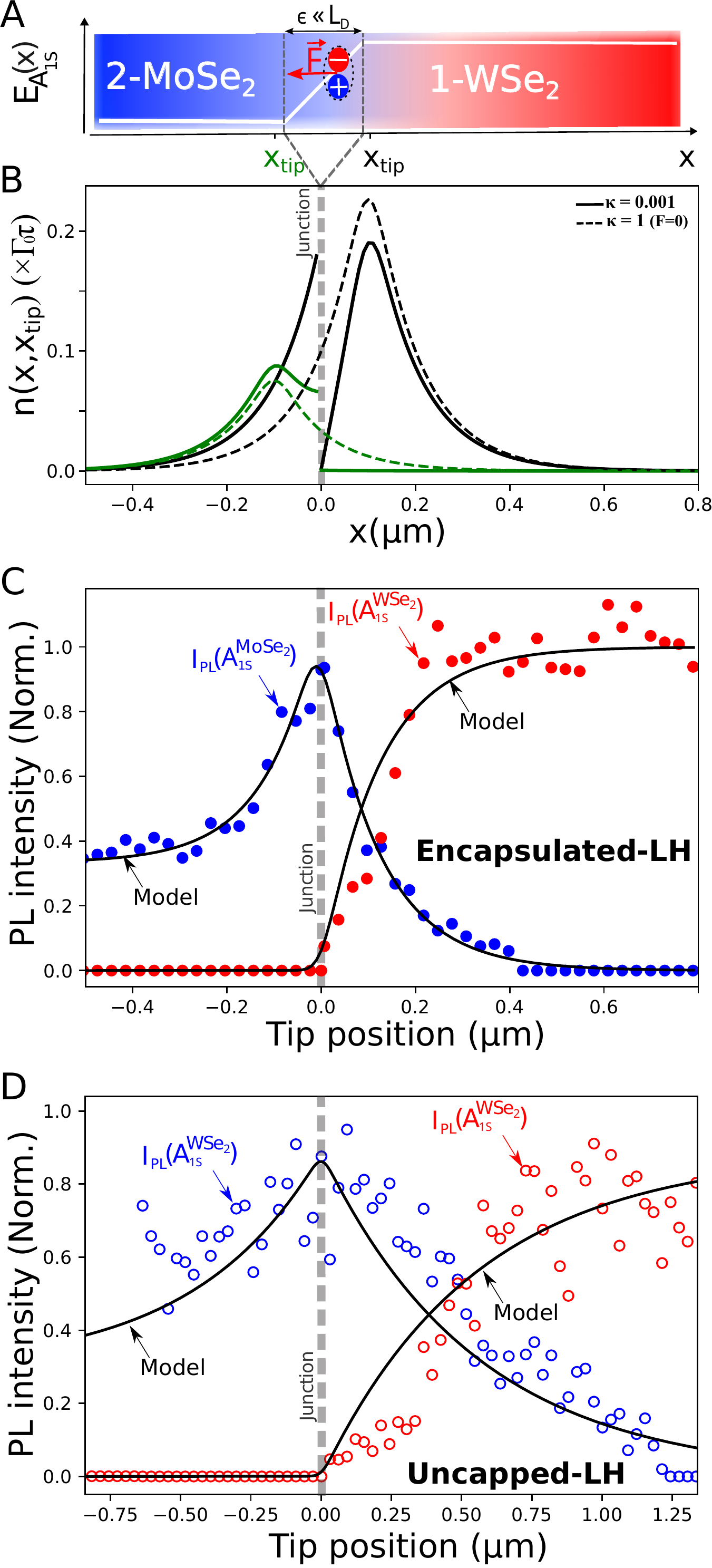} 
	\caption{\textbf{A}. Illustration of the exciton drift inside the interface, the shaded area corresponds to the partition zone between the two materials. \textbf{B}. Exciton density $n(x,x_{tip})$ calculated with the near-field model for two tip positions $x_{tip}=L_{D}$ (black line) and $x_{tip}=-L_{D}$ (green line) with: $\Gamma_{0_1} = 3 \Gamma_{0_2} =\Gamma_{0}$, $\tau_{1}=\tau_{2}=\tau$, $L_{1}=L_{2}=L_{D}=0.1 \mu m$. The dashed lines indicate the exciton density with no junction influence ($F=0$). \textbf{C}. Normalized PL intensity of A$_{1s}^{MoSe_{2}}$ (blue) and A$_{1s}^{WSe_{2}}$ (red) in e-LH area. \textbf{D}. same as \textbf{C} in the un-LH area. The solid black line represents the PL intensities calculated using equations \ref{eq4} and \ref{eq5}.}
	\label{fig3}
\end{figure}

\section{\label{sec:level1} modified exciton transfer model}
In order to get a better understanding of the underlying physical phenomena taking place close to the interface, we developed an exciton transfer model, which is detailed in Supplementary. We theoretically investigate, in the low exciton density regime (typically 10$^{11}$ cm$^{-2}$), the variation of neutral exciton density versus tip position, to be compared to the TEPL results. To do so, we analytically solved the linear 1D steady-state diffusion equation in each material $i$  (1 = WSe$_{2}$, 2 = MoSe$_{2}$), given by :

\begin{equation}
\centering
D_{i}\frac{d^2n(x,x_{tip})}{dx^2} - \frac{n(x,x_{tip})}{\tau_{i}} + \Gamma_{i}(x,x_{tip}) = 0
\end{equation}

$n(x,x_{tip})$ is the exciton density at $x$ position with the excitation taking place at $x_{tip}$, the $x$-axis origin being placed at the LH interface, $D_{i}$ the effective diffusion coefficient and $\tau_{i}$ the effective lifetime, radiative ($\tau_{i}^r$) and non-radiative ($\tau_{i}^{nr}$) (1/$\tau_{i}$ = 1/$\tau_{i}^{r}$+1/$\tau_{i}^{nr}$). $L_{i} = \sqrt{D_{i} \tau_{i}}$ represents the effective diffusion length in the material $i$. We use $\Gamma_{i}(x,x_{tip}) = \Gamma_{0_i}(P,\alpha_{i},\nu) e^{-\frac{(x-x_{tip})^2}{w^2}}$ to simulate the exciton generation under the tip, centered at $x_{tip}$. The width $w$ corresponds to the tip diameter ($\sim 30$nm), $P$ is the enhanced laser power under the tip and $\alpha_{i}$ the absorption coefficient of the material $i$ at the laser energy $h\nu$. We set boundary conditions to be $n(x \rightarrow \pm \infty,x_{tip})=0$. In agreement with experimental results, we impose the continuity of the integrated PL intensity for all values of $x_{tip}$. Finally, we considered dark and bright excitons as non-interacting species. 

As illustrated in Fig.\ref{fig3} \textbf{A}, we model the junction as an ideally thin interface of fixed width $\epsilon \ll L_{D}$, where no electron-hole pair recombination can occur. To model the asymmetric local effective drift of neutral exciton through the junction, we introduce a local uniform force field:

\begin{equation}
\centering
\Vec{F} = - \Vec{\nabla}E_{A_{1s}} \simeq - (E(A_{1s}^{WSe_{2}})-E(A_{1s}^{MoSe_{2}}))/ \epsilon \quad \widehat{x}  
\end{equation}

As a result, the junction imposes the continuity of the excitonic flux density on both sides of the interface by a local constant  flux density $\Vec{j}_{n} = \mu_b \vec{F}n - D_b \vec\nabla{n}$, where $\mu_b$ and $D_{b}$ are the exciton mobility and diffusion coefficient inside the interface.

Fig.\ref{fig3} \textbf{B} displays the calculated $n(x,x_{tip})$ for two different tip positions. We emphasize first on the lateral heterostructure specificity: the exciton density is discontinuous at the interface (see continuous lines). As a reference, the continuous exciton density calculated for classical diffusion ($F=0$) is display in dashed lines. When the tip excitation takes place inside MoSe$_2$, the excitons are blocked by the junction, and the exciton density on the other side of the junction stays close to zero (see continuous green line). Excitons interacting with the junction loose a velocity $v^* = \mu_b F$ blocking the diffusion process. This blocking phenomenon causes a redistribution of the exciton that strongly accumulate at the interface. On the other hand, when the excitation takes place inside WSe$_2$, the excitons cross the junction, and are accelerated, hence acquiring an extra velocity  $v^*$. One can clearly see that this drives those excitons further inside the MoSe$_2$ layer, while bringing the excitonic density down to zero near the interface on the WSe$_{2}$ side (see continuous black line). 

To facilitate further developments, and better characterize the exciton discontinuity at junction we define a partition coefficient linking the exciton densities $n(x=0^{+}$) and $n(x=0^{-}$) on both sides of the interface. This partition coefficient $\kappa = n({x=0^{+}}) \slash n({x=0^{-}}) $ can be written, depending on the excitation position, as:

\begin{equation}
\centering
   \kappa \simeq 
   \left\{
\begin{array}{rcr}
\kappa^+ = \left(1-\frac{L_2}{\mu_b F \tau_{2}}\right)e^{-\frac{\mu_b F \epsilon}{D_b}} + \frac{L_2}{\mu_b F \tau_{2}}  \qquad  x_{tip} > 0 \\
\kappa^- = \left[(1+\frac{L_1}{\mu_b F \tau_{1}})e^{\frac{\mu_b F \epsilon}{D_b}} - \frac{L_1}{\mu_b F \tau_{1}}\right]^{-1} \quad  x_{tip} < 0
\end{array}
\right.   
\end{equation}

$\kappa$ quantifies the discontinuity of the exciton density at the junction, with  $\kappa^{+} = \kappa (x_{tip}>0)$ describing the exciton diffusion from WSe$_2$ to MoSe$_2$  and  $\kappa^{-} = \kappa (x_{tip}<0)$ the diffusion from MoSe$_2$ to WSe$_2$. In our case, as the local equilibrium is established at the steady state, 
 $\kappa^{+}\approx \kappa^{-}$, and to simplify the discussion we will refer simply to $\kappa$ independently of the source position.

When $\kappa = 1$, the exciton distribution, displayed as a dashed line in Fig. \ref{fig3} \textbf{B}, is continuous as observed in classical diffusion. When $\kappa \ll 1 $, corresponding to the present case, the exciton distribution is strongly asymmetric (see continuous line in Fig. \ref{fig3} \textbf{B}). Our finding can be interpreted as an extrapolation of the interfacial thermal resistance so-called Kapitza resistance, which describes the temperature discontinuity at atomically flat interface between two materials  \cite{alosious_kapitza_2020, pollack_kapitza_1969}. For high quality LHs, the absolute value of the exciton Kapitza resistance can be defined as a function of the partition coefficient as follows: (See more details in Supplement).

\begin{equation}
   R_{n} = \frac{n({x=0^{-}})-n({x=0^{+}})}{j_{n}} \simeq \frac{\kappa-1}{v^*}\frac{1-e^{v^*\epsilon/D_{b}}}{1-\kappa e^{v^*\epsilon/D_{b}}}
  \label{resKapitza1}
 \end{equation} 
\vspace{0.3cm}

To compare directly the prediction of our near-field model with the experimental results, we calculate I$_{PL}$(A$_{1s}^{MoSe_{2}}$) and I$_{PL}$(A$_{1s}^{WSe_{2}}$), the integrated PL intensities of each material as a function of the tip position. In the linear regime, the normalized PL intensity of MoSe$_2$ and WSe$_2$  can be written as (see Supplement for details):

\begin{widetext}
\begin{equation}
\centering
I_{PL}^{Norm}(A_{1s}^{MoSe_{2}})(x_{tip})  = \frac{{A}_{\kappa}}{\sqrt{\pi}w}\int_{0}^{\infty}dx  e^{-x/L_{1}}  e^{-\frac{(x-x_{tip})^2}{w^2}} +  \frac{1}{\sqrt{\pi}\beta w}\int_{-\infty}^{0}dx  \left(1+(\beta A_{\kappa}-1) e^{x/L_{2}}\right)e^{-\frac{(x-x_{tip})^2}{w^2}}
\label{eq4}
\end{equation}
\begin{equation}
\centering
I_{PL}^{Norm}(A_{1s}^{WSe_{2}})(x_{tip}) = \frac{1}{\sqrt{\pi} w}\int_{0}^{\infty}dx \left(1-(1-B_{\kappa}) e^{-x/L_{1}} \right) e^{-\frac{(x-x_{tip})^2}{w^2}} +   \frac{B_{\kappa}}{\sqrt{\pi} w}\int_{-\infty}^{0}dx   e^{x/L_{2}}e^{-\frac{(x-x_{tip})^2}{w^2}}
\label{eq5}
\end{equation}
\end{widetext}
with $I_{PL}^\infty(A_{1s}^{WSe_{2}})$ used for normalization. The experimental PL is fitted using four fitting parameters: ${A}_{\kappa}$ and ${B}_{\kappa}$, amplitude parameters, giving us access to $\kappa$, the partition coefficient  and $L_1$, $L_2$ the diffusion lengths of both materials. Fig.\ref{fig3} \textbf{C} and \textbf{D}, show the experimental normalized PL intensities of MoSe$_2$ (blue dots) and WSe$_2$ (red dots) for both e-LH and un-LH configurations respectively. The black solid lines represent the fits using equations \ref{eq4} and \ref{eq5}. The resulting fitting parameters are displayed in table \ref{fit-table}.  In both systems, the $\kappa \ll 1$  values for e-LH and un-LH confirm the strong asymmetry of the junction and the resulting strong Kapitza exciton resistance.
\begin{table}[h]
\centering
\begin{tabular}{|c|c|c|c|c|}
\hline
\textbf{Area} & {$L_{1} (nm)$} & {$L_{2} (nm)$} & {$\kappa$} & R$_n$(s.m$^{-1}$)  \\
\hline
e-LH   & $120 \pm 6$ & $110 \pm 5.5$ &$\sim 1.10^{-3}$& $\sim 1.10^{-5}$ \\
un-LH   & $550 \pm 55$ &$450 \pm 45$ & $\sim 1.10^{-3}$ & $\sim 1.10^{-5}$  \\
\hline
\end{tabular}
\caption{\label{fit-table}Results of the PL fits using the near-field model for both e-LH and un-LH systems.}
\end{table}
\section{\label{sec:level1} DISCUSSION}
One can see that the model describes extremely well the strong quenching of the WSe$_2$-related PL and the appearance of MoSe$_2$-related PL when approaching the junction from the right ($x_{tip}>0$), shedding light on the experimental results on unidirectional excitonic transport across the junction. It is also able to quantitatively describe the enhancement of $I_{PL}$(A$_{1s}^{MoSe_{2}}$) at the junction and its progressive decrease to the typically lower value $I_{PL}^\infty$(A$_{1s}^{MoSe_{2}}$) of monolayer MoSe$_2$  as the tip is moved away from the junction ($x_{tip}<0$). We would like to point out that this observation is far from trivial, it indicates that to conserve the experimentally observed continuity of the PL intensity, the exciton density at the interface ($n(x=0^{\pm})$) is enhanced. We believe that this could be linked to the drastic change in the exciton velocity at the interface. Indeed, when the tip is inside MoSe$_2$, excitons diffusing toward WSe$_2$ are abruptly stopped at the junction, their average velocity going to zero. The non-radiative lifetime  ($\tau_{2}^{nr}$), being in part linked to the probability for the exciton to encounter non-radiative traps during its lifetime, is by consequence increased \cite{zipfel_exciton_2020}, resulting in an increased population at the interface. This would explain why the size of PL enhancement area (where $I_{PL}$(A$_{1s}^{MoSe_{2}}$) $\ge$ $I_{PL}^\infty$(A$_{1s}^{MoSe_{2}}$)) is a function of the diffusion length (see Figure \ref{fig3}), as it only occurs when excitons start accumulating at the junction.

\begin{figure}
	\includegraphics[width=9cm]{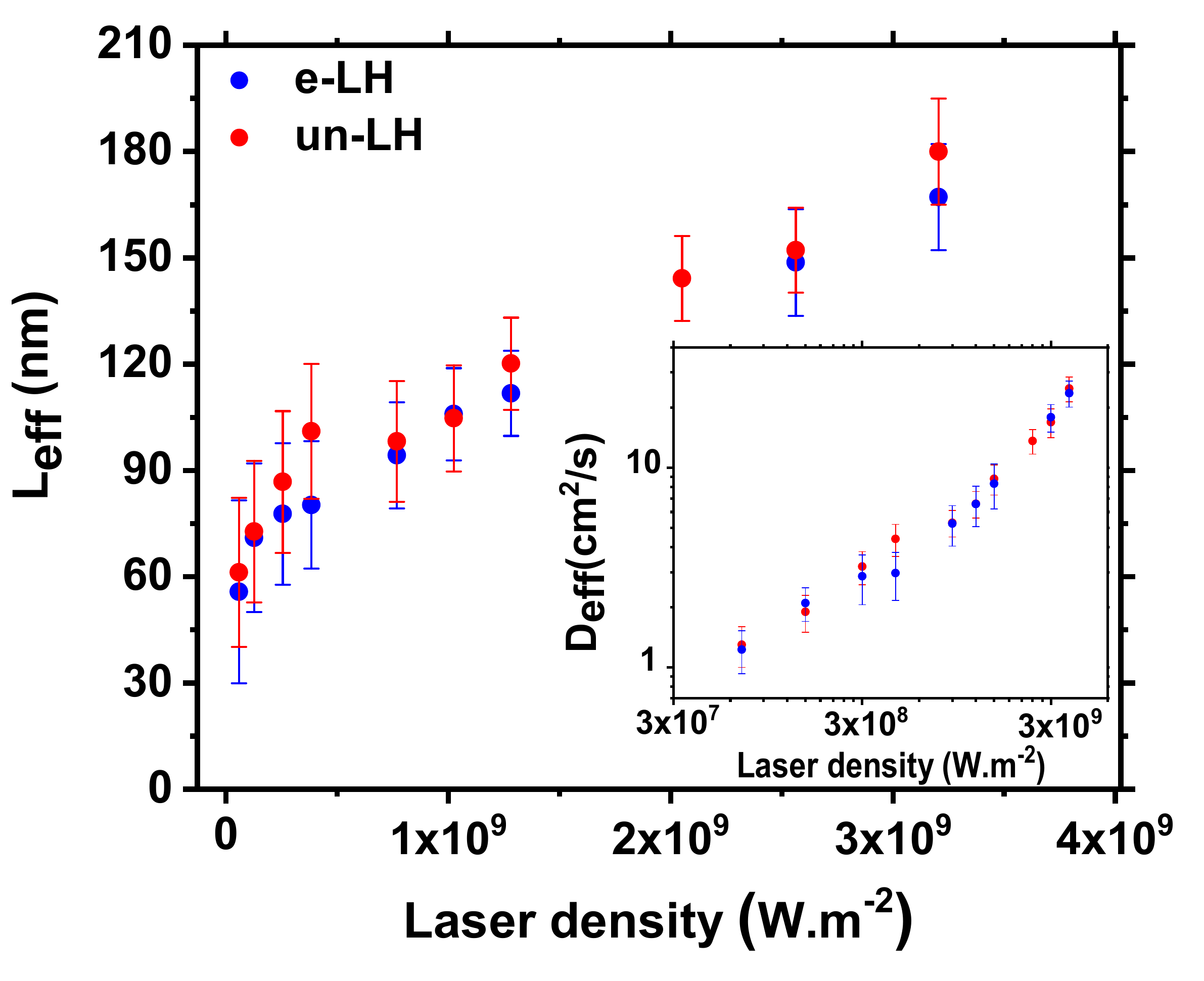} 
	\caption{Evolution of the effective diffusion length  (L$_{eff}$) with the excitation power density measured in $WSe_2$. Results obtained from time revolved PL profiles (see supplement). The inset shows the power dependence of the effective diffusion coefficient (D$_{eff}$).} 
	\label{fig1b}
\end{figure}

Finally, the model reveals that the important difference observed in the PL profile between e-LH and un-LH is linked to drastically different values of the diffusion length, much shorter in e-LH than in un-LH (see experimental results in Fig. \ref{fig2} and theoretical results in table \ref{fit-table}). To understand why $L_D^{e-LH} < L_D^{un-LH}$ we need to consider that even if the TEPL experiments on e-LH and un-LH were performed using the same laser power (400$\mu$W), the presence or absence of the top hBN layer, strongly impacts the near-field configuration under the tip. In first approximation, and considering a spherical tip of 30nm diameter,  we can estimate the near-field intensity ratio on the TMD layer as:
\begin{equation}
\centering 
\dfrac{\left\| \Vec{E}\right\|_{un-LH}^{2}}{\left\|\Vec{E}\right\|_{e-LH}^{2}} \simeq \left(\frac{Z_{e-LH}}{Z_{un-LH}}\right)^{6}\left(\frac{n_{BN}}{n_{air}}\right)^{4}
\label{eq7}
\end{equation}
$Z_{e-LH}$ and $Z_{un-LH}$ are the distances from the tip center to the TMD-ML in e-LH and un-LH areas, respectively. $n_{BN}$ and $n_{air}$ are hBN and air refractive index, respectively. Considering a 3 nm-thick hBN top layer (measured by AFM), we can estimate the enhancement ratio to be $\sim 23$. We can then expect the exciton density generated under the tip in un-LH to be more than one order of magnitude larger than in e-LH. To characterize the exciton transport properties of the WSe$_{2}$ layer versus exciton density we use spatio-temporal PL signal in a confocal microscope. A local pulsed laser excitation ($\lambda$= 692 nm, pulse duration 1.5 ps), generates excitons, and we record the time evolution of the PL profile. We then extract the time evolution of the squared width $w^2(t)=w^2(0)+\Delta w^2(t)$ which is used to determine an effective diffusion coefficient in the 80 first picosecond where 90\% of the PL signal originates : $\Delta w^2(t)=4D_{eff} t$. An important point is to be able to decorrelate the effective lifetime $\tau_{eff}$ (linked to the PL decay) and the diffusion coefficients, from which we deduce the effective diffusion length $L_{eff}=\sqrt{D_{eff}\tau_{eff}}$. Experimental conditions and detailed results are given in supplements. The excitation power density was varied over several order of magnitude and the results compared between the e-LH and un-LH samples. Fig. \ref{fig1b} shows that, for both samples, $L_{eff}$ increases with the excitation power density (i.e. the excitonic dentity) which is mainly due to the increase of the effective diffusion coefficient (See Supplement). This trend, attributed to a weak Auger contribution has been previously observed in WS$_2$ \cite{uddin_enhanced_2022, zipfel_exciton_2020} and explain the large difference observed in the diffusion lengths we observed in e-LH and un-LH. The experimental results have been modeled by using a Auger coefficient of $0.2 ~10^{-6}$cm$^2$/s (See Supplementary). This value is two orders of magnitude lower than the one estimated on non encapsulated WS$_2$ (without top and bottom hBN) where a large difference of the transport properties values with a fully encapsulated layer was observed \citep{zipfel_exciton_2020}. In summary, we show that both samples exhibit similar exciton transport properties, indicating the importance of the hBN bottom layer that prevents strong Auger scattering effect.

One can notice here that, according to Equation \ref{eq7}, the enhancement ratio is dominated by the tip-TMD layer distance. This offers a unique opportunity, as using plasmonic tips or nanoantennas in combination with hBN encapsulation allows to control the diffusion length in this type of structures. As a matter of fact, controlling the thickness of the top hBN layer from a single layer to 20 layers, for example, would change the enhancement factor by a factor 10, modifying the diffusion length by the same amount.

\vspace{0.3cm}

In summary, we have performed a detailed tip-enhanced spectroscopy study of a MoSe$_{2}$-WSe$_{2}$ lateral heterostructure, and have developed a model to render the observation of the unilateral transport across the junction observed in near-field PL experiment. It accounts for a discontinuity of the exciton density at the interface. We have thus shown that the exciton diffusion properties follow a semi-classical process, due to the difference in the energy gap between the two materials: the usual isotropic in-plane diffusion of the excitons is frustrated by the junction, leading to an asymmetric diffusion, in which all generated excitons move away from the high bandgap WSe$_2$ layer to recombine in the low bandgap MoSe$_2$ layer. This transfer causes near the interface the quenching of WSe$_2$-related PL and the enhancement of the MoSe$_2$-related PL, well above the one observed in "bulk" MoSe$_2$ far from the junction. Interestingly, we observe similar asymmetric diffusion property for samples with or without top hBN. Both samples present similar intrinsic transport properties, probably linked to an efficient screening of the dielectric disorder by the bottom encapsulation by hBN layer.
Finally, we have shown that the diffusion length in WSe$_2$ is strongly dependent on the exciton density. This offers a new degree of freedom, as changing the laser power density or the near-field enhancement (for instance using optically resonant nanoantennas instead of the plasmonic tip) would allow tuning the diffusion length to any wanted value from tens of nanometer up to few micrometers. TMD-based lateral heterostructure is a rapidly evolving research field that could offer to combine TMDs with very different bandgaps, thus allowing  partition zone engineering, or creating more complex designs using three or more TMDs (lateral excitonic quantum wells). This work offers both the theoretical and experimental tools to predict and control the new diffusion properties that will be at the origin of new excitronic devices.

\begin{acknowledgments}
Toulouse acknowledges partial funding from ANR IXTASE, ANR HiLight, ANR Ti-P, NanoX project 2DLight, the Institut Universitaire de France, and the EUR grant ATRAP-2D NanoX ANR-17-EURE-0009 in the framework of the “Programme des Investissements d’Avenir”, the Institute of quantum technology in Occitanie IQO and a UPS excellence PhD grant. Growth of hexagonal boron nitride crystals was supported by JSPS KAKENHI (Grants No. 19H05790, No. 20H00354 and No. 21H05233). The Jena group received financial support of the Deutsche Forschungsgemeinschaft (DFG) through a research infrastructure grant INST 275/257-1 FUGG, CRC 1375 NOA (Project B2), SPP2244 (Project TU149/13-1) as well as DFG grant TU149/16-1. This project has also received funding from the joint European Union’s Horizon 2020 and DFG research and innovation program FLAG-ERA under grant TU149/9-1.
\dots.
\end{acknowledgments}

\bibliography{heterostructureLaterale2}

\begin{thebibliography}{26}%
\makeatletter
\providecommand \@ifxundefined [1]{%
 \@ifx{#1\undefined}
}%
\providecommand \@ifnum [1]{%
 \ifnum #1\expandafter \@firstoftwo
 \else \expandafter \@secondoftwo
 \fi
}%
\providecommand \@ifx [1]{%
 \ifx #1\expandafter \@firstoftwo
 \else \expandafter \@secondoftwo
 \fi
}%
\providecommand \natexlab [1]{#1}%
\providecommand \enquote  [1]{``#1''}%
\providecommand \bibnamefont  [1]{#1}%
\providecommand \bibfnamefont [1]{#1}%
\providecommand \citenamefont [1]{#1}%
\providecommand \href@noop [0]{\@secondoftwo}%
\providecommand \href [0]{\begingroup \@sanitize@url \@href}%
\providecommand \@href[1]{\@@startlink{#1}\@@href}%
\providecommand \@@href[1]{\endgroup#1\@@endlink}%
\providecommand \@sanitize@url [0]{\catcode `\\12\catcode `\$12\catcode
  `\&12\catcode `\#12\catcode `\^12\catcode `\_12\catcode `\%12\relax}%
\providecommand \@@startlink[1]{}%
\providecommand \@@endlink[0]{}%
\providecommand \url  [0]{\begingroup\@sanitize@url \@url }%
\providecommand \@url [1]{\endgroup\@href {#1}{\urlprefix }}%
\providecommand \urlprefix  [0]{URL }%
\providecommand \Eprint [0]{\href }%
\providecommand \doibase [0]{https://doi.org/}%
\providecommand \selectlanguage [0]{\@gobble}%
\providecommand \bibinfo  [0]{\@secondoftwo}%
\providecommand \bibfield  [0]{\@secondoftwo}%
\providecommand \translation [1]{[#1]}%
\providecommand \BibitemOpen [0]{}%
\providecommand \bibitemStop [0]{}%
\providecommand \bibitemNoStop [0]{.\EOS\space}%
\providecommand \EOS [0]{\spacefactor3000\relax}%
\providecommand \BibitemShut  [1]{\csname bibitem#1\endcsname}%
\let\auto@bib@innerbib\@empty
\bibitem [{\citenamefont {Butov}(2017)}]{butov_excitonic_2017}%
  \BibitemOpen
  \bibfield  {author} {\bibinfo {author} {\bibfnamefont {L.}~\bibnamefont
  {Butov}},\ }\bibfield  {title} {\bibinfo {title} {Excitonic devices},\ }\href
  {https://doi.org/10.1016/j.spmi.2016.12.035} {\bibfield  {journal} {\bibinfo
  {journal} {Superlattices and Microstructures}\ }\textbf {\bibinfo {volume}
  {108}},\ \bibinfo {pages} {2} (\bibinfo {year} {2017})}\BibitemShut {NoStop}%
\bibitem [{\citenamefont {Miller}(2000)}]{867687}%
  \BibitemOpen
  \bibfield  {author} {\bibinfo {author} {\bibfnamefont {D.}~\bibnamefont
  {Miller}},\ }\bibfield  {title} {\bibinfo {title} {Rationale and challenges
  for optical interconnects to electronic chips},\ }\href
  {https://doi.org/10.1109/5.867687} {\bibfield  {journal} {\bibinfo  {journal}
  {Proceedings of the IEEE}\ }\textbf {\bibinfo {volume} {88}},\ \bibinfo
  {pages} {728} (\bibinfo {year} {2000})}\BibitemShut {NoStop}%
\bibitem [{\citenamefont {High}\ \emph {et~al.}(2008)\citenamefont {High},
  \citenamefont {Novitskaya}, \citenamefont {Butov}, \citenamefont {Hanson},\
  and\ \citenamefont {Gossard}}]{high_control_2008}%
  \BibitemOpen
  \bibfield  {author} {\bibinfo {author} {\bibfnamefont {A.~A.}\ \bibnamefont
  {High}}, \bibinfo {author} {\bibfnamefont {E.~E.}\ \bibnamefont
  {Novitskaya}}, \bibinfo {author} {\bibfnamefont {L.~V.}\ \bibnamefont
  {Butov}}, \bibinfo {author} {\bibfnamefont {M.}~\bibnamefont {Hanson}},\ and\
  \bibinfo {author} {\bibfnamefont {A.~C.}\ \bibnamefont {Gossard}},\
  }\bibfield  {title} {\bibinfo {title} {Control of {Exciton} {Fluxes} in an
  {Excitonic} {Integrated} {Circuit}},\ }\href
  {https://doi.org/10.1126/science.1157845} {\bibfield  {journal} {\bibinfo
  {journal} {Science}\ }\textbf {\bibinfo {volume} {321}},\ \bibinfo {pages}
  {229} (\bibinfo {year} {2008})},\ \bibinfo {note} {\_eprint:
  https://www.science.org/doi/pdf/10.1126/science.1157845}\BibitemShut
  {NoStop}%
\bibitem [{\citenamefont {Peng}\ \emph {et~al.}(2022)\citenamefont {Peng},
  \citenamefont {Ripin}, \citenamefont {Ye}, \citenamefont {Zhu}, \citenamefont
  {Wu}, \citenamefont {Lee}, \citenamefont {Li}, \citenamefont {Taniguchi},
  \citenamefont {Watanabe}, \citenamefont {Cao}, \citenamefont {Xu},\ and\
  \citenamefont {Li}}]{peng_long-range_2022}%
  \BibitemOpen
  \bibfield  {author} {\bibinfo {author} {\bibfnamefont {R.}~\bibnamefont
  {Peng}}, \bibinfo {author} {\bibfnamefont {A.}~\bibnamefont {Ripin}},
  \bibinfo {author} {\bibfnamefont {Y.}~\bibnamefont {Ye}}, \bibinfo {author}
  {\bibfnamefont {J.}~\bibnamefont {Zhu}}, \bibinfo {author} {\bibfnamefont
  {C.}~\bibnamefont {Wu}}, \bibinfo {author} {\bibfnamefont {S.}~\bibnamefont
  {Lee}}, \bibinfo {author} {\bibfnamefont {H.}~\bibnamefont {Li}}, \bibinfo
  {author} {\bibfnamefont {T.}~\bibnamefont {Taniguchi}}, \bibinfo {author}
  {\bibfnamefont {K.}~\bibnamefont {Watanabe}}, \bibinfo {author}
  {\bibfnamefont {T.}~\bibnamefont {Cao}}, \bibinfo {author} {\bibfnamefont
  {X.}~\bibnamefont {Xu}},\ and\ \bibinfo {author} {\bibfnamefont
  {M.}~\bibnamefont {Li}},\ }\bibfield  {title} {\bibinfo {title} {Long-range
  transport of {2D} excitons with acoustic waves},\ }\href
  {https://doi.org/10.1038/s41467-022-29042-9} {\bibfield  {journal} {\bibinfo
  {journal} {Nature Communications}\ }\textbf {\bibinfo {volume} {13}},\
  \bibinfo {pages} {1334} (\bibinfo {year} {2022})}\BibitemShut {NoStop}%
\bibitem [{\citenamefont {High}\ \emph {et~al.}(2007)\citenamefont {High},
  \citenamefont {Hammack}, \citenamefont {Butov}, \citenamefont {Hanson},\ and\
  \citenamefont {Gossard}}]{high_exciton_2007}%
  \BibitemOpen
  \bibfield  {author} {\bibinfo {author} {\bibfnamefont {A.~A.}\ \bibnamefont
  {High}}, \bibinfo {author} {\bibfnamefont {A.~T.}\ \bibnamefont {Hammack}},
  \bibinfo {author} {\bibfnamefont {L.~V.}\ \bibnamefont {Butov}}, \bibinfo
  {author} {\bibfnamefont {M.}~\bibnamefont {Hanson}},\ and\ \bibinfo {author}
  {\bibfnamefont {A.~C.}\ \bibnamefont {Gossard}},\ }\bibfield  {title}
  {\bibinfo {title} {Exciton optoelectronic transistor},\ }\href
  {https://doi.org/10.1364/OL.32.002466} {\bibfield  {journal} {\bibinfo
  {journal} {Optics Letters}\ }\textbf {\bibinfo {volume} {32}},\ \bibinfo
  {pages} {2466} (\bibinfo {year} {2007})}\BibitemShut {NoStop}%
\bibitem [{\citenamefont {Wang}\ \emph
  {et~al.}(2018{\natexlab{a}})\citenamefont {Wang}, \citenamefont {Chernikov},
  \citenamefont {Glazov}, \citenamefont {Heinz}, \citenamefont {Marie},
  \citenamefont {Amand},\ and\ \citenamefont
  {Urbaszek}}]{wang_colloquium_2018}%
  \BibitemOpen
  \bibfield  {author} {\bibinfo {author} {\bibfnamefont {G.}~\bibnamefont
  {Wang}}, \bibinfo {author} {\bibfnamefont {A.}~\bibnamefont {Chernikov}},
  \bibinfo {author} {\bibfnamefont {M.~M.}\ \bibnamefont {Glazov}}, \bibinfo
  {author} {\bibfnamefont {T.~F.}\ \bibnamefont {Heinz}}, \bibinfo {author}
  {\bibfnamefont {X.}~\bibnamefont {Marie}}, \bibinfo {author} {\bibfnamefont
  {T.}~\bibnamefont {Amand}},\ and\ \bibinfo {author} {\bibfnamefont
  {B.}~\bibnamefont {Urbaszek}},\ }\bibfield  {title} {\bibinfo {title}
  {\textit{{Colloquium}} : {Excitons} in atomically thin transition metal
  dichalcogenides},\ }\href {https://doi.org/10.1103/RevModPhys.90.021001}
  {\bibfield  {journal} {\bibinfo  {journal} {Reviews of Modern Physics}\
  }\textbf {\bibinfo {volume} {90}},\ \bibinfo {pages} {021001} (\bibinfo
  {year} {2018}{\natexlab{a}})}\BibitemShut {NoStop}%
\bibitem [{\citenamefont {Wang}\ \emph
  {et~al.}(2018{\natexlab{b}})\citenamefont {Wang}, \citenamefont
  {Verzhbitskiy},\ and\ \citenamefont {Eda}}]{wang_electroluminescent_2018}%
  \BibitemOpen
  \bibfield  {author} {\bibinfo {author} {\bibfnamefont {J.}~\bibnamefont
  {Wang}}, \bibinfo {author} {\bibfnamefont {I.}~\bibnamefont {Verzhbitskiy}},\
  and\ \bibinfo {author} {\bibfnamefont {G.}~\bibnamefont {Eda}},\ }\bibfield
  {title} {\bibinfo {title} {Electroluminescent {Devices} {Based} on {2D}
  {Semiconducting} {Transition} {Metal} {Dichalcogenides}},\ }\href
  {https://doi.org/10.1002/adma.201802687} {\bibfield  {journal} {\bibinfo
  {journal} {Advanced Materials}\ }\textbf {\bibinfo {volume} {30}},\ \bibinfo
  {pages} {1802687} (\bibinfo {year} {2018}{\natexlab{b}})}\BibitemShut
  {NoStop}%
\bibitem [{\citenamefont {Novoselov}\ \emph {et~al.}(2016)\citenamefont
  {Novoselov}, \citenamefont {Mishchenko}, \citenamefont {Carvalho},\ and\
  \citenamefont {Castro~Neto}}]{novoselov_2d_2016}%
  \BibitemOpen
  \bibfield  {author} {\bibinfo {author} {\bibfnamefont {K.~S.}\ \bibnamefont
  {Novoselov}}, \bibinfo {author} {\bibfnamefont {A.}~\bibnamefont
  {Mishchenko}}, \bibinfo {author} {\bibfnamefont {A.}~\bibnamefont
  {Carvalho}},\ and\ \bibinfo {author} {\bibfnamefont {A.~H.}\ \bibnamefont
  {Castro~Neto}},\ }\bibfield  {title} {\bibinfo {title} {{2D} materials and
  van der {Waals} heterostructures},\ }\href
  {https://doi.org/10.1126/science.aac9439} {\bibfield  {journal} {\bibinfo
  {journal} {Science}\ }\textbf {\bibinfo {volume} {353}},\ \bibinfo {pages}
  {aac9439} (\bibinfo {year} {2016})}\BibitemShut {NoStop}%
\bibitem [{\citenamefont {Cheiwchanchamnangij}\ and\ \citenamefont
  {Lambrecht}(2012)}]{cheiwchanchamnangij_quasiparticle_2012}%
  \BibitemOpen
  \bibfield  {author} {\bibinfo {author} {\bibfnamefont {T.}~\bibnamefont
  {Cheiwchanchamnangij}}\ and\ \bibinfo {author} {\bibfnamefont {W.~R.~L.}\
  \bibnamefont {Lambrecht}},\ }\bibfield  {title} {\bibinfo {title}
  {Quasiparticle band structure calculation of monolayer, bilayer, and bulk
  {MoS} 2},\ }\href {https://doi.org/10.1103/PhysRevB.85.205302} {\bibfield
  {journal} {\bibinfo  {journal} {Physical Review B}\ }\textbf {\bibinfo
  {volume} {85}},\ \bibinfo {pages} {205302} (\bibinfo {year}
  {2012})}\BibitemShut {NoStop}%
\bibitem [{\citenamefont {Mak}\ \emph {et~al.}(2014)\citenamefont {Mak},
  \citenamefont {McGill}, \citenamefont {Park},\ and\ \citenamefont
  {McEuen}}]{mak_valley_2014}%
  \BibitemOpen
  \bibfield  {author} {\bibinfo {author} {\bibfnamefont {K.~F.}\ \bibnamefont
  {Mak}}, \bibinfo {author} {\bibfnamefont {K.~L.}\ \bibnamefont {McGill}},
  \bibinfo {author} {\bibfnamefont {J.}~\bibnamefont {Park}},\ and\ \bibinfo
  {author} {\bibfnamefont {P.~L.}\ \bibnamefont {McEuen}},\ }\bibfield  {title}
  {\bibinfo {title} {The valley {Hall} effect in {MoS} $_{\textrm{2}}$
  transistors},\ }\href {https://doi.org/10.1126/science.1250140} {\bibfield
  {journal} {\bibinfo  {journal} {Science}\ }\textbf {\bibinfo {volume}
  {344}},\ \bibinfo {pages} {1489} (\bibinfo {year} {2014})}\BibitemShut
  {NoStop}%
\bibitem [{\citenamefont {Ubrig}\ \emph {et~al.}(2017)\citenamefont {Ubrig},
  \citenamefont {Jo}, \citenamefont {Philippi}, \citenamefont {Costanzo},
  \citenamefont {Berger}, \citenamefont {Kuzmenko},\ and\ \citenamefont
  {Morpurgo}}]{ubrig_microscopic_2017}%
  \BibitemOpen
  \bibfield  {author} {\bibinfo {author} {\bibfnamefont {N.}~\bibnamefont
  {Ubrig}}, \bibinfo {author} {\bibfnamefont {S.}~\bibnamefont {Jo}}, \bibinfo
  {author} {\bibfnamefont {M.}~\bibnamefont {Philippi}}, \bibinfo {author}
  {\bibfnamefont {D.}~\bibnamefont {Costanzo}}, \bibinfo {author}
  {\bibfnamefont {H.}~\bibnamefont {Berger}}, \bibinfo {author} {\bibfnamefont
  {A.~B.}\ \bibnamefont {Kuzmenko}},\ and\ \bibinfo {author} {\bibfnamefont
  {A.~F.}\ \bibnamefont {Morpurgo}},\ }\bibfield  {title} {\bibinfo {title}
  {Microscopic {Origin} of the {Valley} {Hall} {Effect} in {Transition} {Metal}
  {Dichalcogenides} {Revealed} by {Wavelength}-{Dependent} {Mapping}},\ }\href
  {https://doi.org/10.1021/acs.nanolett.7b02666} {\bibfield  {journal}
  {\bibinfo  {journal} {Nano Letters}\ }\textbf {\bibinfo {volume} {17}},\
  \bibinfo {pages} {5719} (\bibinfo {year} {2017})}\BibitemShut {NoStop}%
\bibitem [{\citenamefont {Zhou}\ \emph {et~al.}(2019)\citenamefont {Zhou},
  \citenamefont {Taguchi}, \citenamefont {Kawaguchi}, \citenamefont {Tanaka},\
  and\ \citenamefont {Law}}]{zhou_spin-orbit_2019}%
  \BibitemOpen
  \bibfield  {author} {\bibinfo {author} {\bibfnamefont {B.~T.}\ \bibnamefont
  {Zhou}}, \bibinfo {author} {\bibfnamefont {K.}~\bibnamefont {Taguchi}},
  \bibinfo {author} {\bibfnamefont {Y.}~\bibnamefont {Kawaguchi}}, \bibinfo
  {author} {\bibfnamefont {Y.}~\bibnamefont {Tanaka}},\ and\ \bibinfo {author}
  {\bibfnamefont {K.~T.}\ \bibnamefont {Law}},\ }\bibfield  {title} {\bibinfo
  {title} {Spin-orbit coupling induced valley {Hall} effects in
  transition-metal dichalcogenides},\ }\href
  {https://doi.org/10.1038/s42005-019-0127-7} {\bibfield  {journal} {\bibinfo
  {journal} {Communications Physics}\ }\textbf {\bibinfo {volume} {2}},\
  \bibinfo {pages} {26} (\bibinfo {year} {2019})}\BibitemShut {NoStop}%
\bibitem [{\citenamefont {Kulig}\ \emph {et~al.}(2018)\citenamefont {Kulig},
  \citenamefont {Zipfel}, \citenamefont {Nagler}, \citenamefont {Blanter},
  \citenamefont {Schüller}, \citenamefont {Korn}, \citenamefont {Paradiso},
  \citenamefont {Glazov},\ and\ \citenamefont
  {Chernikov}}]{kulig_exciton_2018}%
  \BibitemOpen
  \bibfield  {author} {\bibinfo {author} {\bibfnamefont {M.}~\bibnamefont
  {Kulig}}, \bibinfo {author} {\bibfnamefont {J.}~\bibnamefont {Zipfel}},
  \bibinfo {author} {\bibfnamefont {P.}~\bibnamefont {Nagler}}, \bibinfo
  {author} {\bibfnamefont {S.}~\bibnamefont {Blanter}}, \bibinfo {author}
  {\bibfnamefont {C.}~\bibnamefont {Schüller}}, \bibinfo {author}
  {\bibfnamefont {T.}~\bibnamefont {Korn}}, \bibinfo {author} {\bibfnamefont
  {N.}~\bibnamefont {Paradiso}}, \bibinfo {author} {\bibfnamefont {M.~M.}\
  \bibnamefont {Glazov}},\ and\ \bibinfo {author} {\bibfnamefont
  {A.}~\bibnamefont {Chernikov}},\ }\bibfield  {title} {\bibinfo {title}
  {Exciton {Diffusion} and {Halo} {Effects} in {Monolayer} {Semiconductors}},\
  }\href {https://doi.org/10.1103/PhysRevLett.120.207401} {\bibfield  {journal}
  {\bibinfo  {journal} {Physical Review Letters}\ }\textbf {\bibinfo {volume}
  {120}},\ \bibinfo {pages} {207401} (\bibinfo {year} {2018})}\BibitemShut
  {NoStop}%
\bibitem [{\citenamefont {Rosati}\ \emph {et~al.}(2020)\citenamefont {Rosati},
  \citenamefont {Perea-Causín}, \citenamefont {Brem},\ and\ \citenamefont
  {Malic}}]{rosati_negative_2020}%
  \BibitemOpen
  \bibfield  {author} {\bibinfo {author} {\bibfnamefont {R.}~\bibnamefont
  {Rosati}}, \bibinfo {author} {\bibfnamefont {R.}~\bibnamefont
  {Perea-Causín}}, \bibinfo {author} {\bibfnamefont {S.}~\bibnamefont
  {Brem}},\ and\ \bibinfo {author} {\bibfnamefont {E.}~\bibnamefont {Malic}},\
  }\bibfield  {title} {\bibinfo {title} {Negative effective excitonic diffusion
  in monolayer transition metal dichalcogenides},\ }\href
  {https://doi.org/10.1039/C9NR07056G} {\bibfield  {journal} {\bibinfo
  {journal} {Nanoscale}\ }\textbf {\bibinfo {volume} {12}},\ \bibinfo {pages}
  {356} (\bibinfo {year} {2020})}\BibitemShut {NoStop}%
\bibitem [{\citenamefont {Unuchek}\ \emph {et~al.}(2018)\citenamefont
  {Unuchek}, \citenamefont {Ciarrocchi}, \citenamefont {Avsar}, \citenamefont
  {Watanabe}, \citenamefont {Taniguchi},\ and\ \citenamefont
  {Kis}}]{unuchek_room-temperature_2018}%
  \BibitemOpen
  \bibfield  {author} {\bibinfo {author} {\bibfnamefont {D.}~\bibnamefont
  {Unuchek}}, \bibinfo {author} {\bibfnamefont {A.}~\bibnamefont {Ciarrocchi}},
  \bibinfo {author} {\bibfnamefont {A.}~\bibnamefont {Avsar}}, \bibinfo
  {author} {\bibfnamefont {K.}~\bibnamefont {Watanabe}}, \bibinfo {author}
  {\bibfnamefont {T.}~\bibnamefont {Taniguchi}},\ and\ \bibinfo {author}
  {\bibfnamefont {A.}~\bibnamefont {Kis}},\ }\bibfield  {title} {\bibinfo
  {title} {Room-temperature electrical control of exciton flux in a van der
  {Waals} heterostructure},\ }\href {https://doi.org/10.1038/s41586-018-0357-y}
  {\bibfield  {journal} {\bibinfo  {journal} {Nature}\ }\textbf {\bibinfo
  {volume} {560}},\ \bibinfo {pages} {340} (\bibinfo {year}
  {2018})}\BibitemShut {NoStop}%
\bibitem [{\citenamefont {Beret}\ \emph {et~al.}(2022)\citenamefont {Beret},
  \citenamefont {Paradisanos}, \citenamefont {Lamsaadi}, \citenamefont {Gan},
  \citenamefont {Najafidehaghani}, \citenamefont {George}, \citenamefont
  {Lehnert}, \citenamefont {Biskupek}, \citenamefont {Kaiser}, \citenamefont
  {Shree}, \citenamefont {Estrada-Real}, \citenamefont {Lagarde}, \citenamefont
  {Marie}, \citenamefont {Renucci}, \citenamefont {Watanabe}, \citenamefont
  {Taniguchi}, \citenamefont {Weber}, \citenamefont {Paillard}, \citenamefont
  {Lombez}, \citenamefont {Poumirol}, \citenamefont {Turchanin},\ and\
  \citenamefont {Urbaszek}}]{beret_exciton_2022}%
  \BibitemOpen
  \bibfield  {author} {\bibinfo {author} {\bibfnamefont {D.}~\bibnamefont
  {Beret}}, \bibinfo {author} {\bibfnamefont {I.}~\bibnamefont {Paradisanos}},
  \bibinfo {author} {\bibfnamefont {H.}~\bibnamefont {Lamsaadi}}, \bibinfo
  {author} {\bibfnamefont {Z.}~\bibnamefont {Gan}}, \bibinfo {author}
  {\bibfnamefont {E.}~\bibnamefont {Najafidehaghani}}, \bibinfo {author}
  {\bibfnamefont {A.}~\bibnamefont {George}}, \bibinfo {author} {\bibfnamefont
  {T.}~\bibnamefont {Lehnert}}, \bibinfo {author} {\bibfnamefont
  {J.}~\bibnamefont {Biskupek}}, \bibinfo {author} {\bibfnamefont
  {U.}~\bibnamefont {Kaiser}}, \bibinfo {author} {\bibfnamefont
  {S.}~\bibnamefont {Shree}}, \bibinfo {author} {\bibfnamefont
  {A.}~\bibnamefont {Estrada-Real}}, \bibinfo {author} {\bibfnamefont
  {D.}~\bibnamefont {Lagarde}}, \bibinfo {author} {\bibfnamefont
  {X.}~\bibnamefont {Marie}}, \bibinfo {author} {\bibfnamefont
  {P.}~\bibnamefont {Renucci}}, \bibinfo {author} {\bibfnamefont
  {K.}~\bibnamefont {Watanabe}}, \bibinfo {author} {\bibfnamefont
  {T.}~\bibnamefont {Taniguchi}}, \bibinfo {author} {\bibfnamefont
  {S.}~\bibnamefont {Weber}}, \bibinfo {author} {\bibfnamefont
  {V.}~\bibnamefont {Paillard}}, \bibinfo {author} {\bibfnamefont
  {L.}~\bibnamefont {Lombez}}, \bibinfo {author} {\bibfnamefont {J.-M.}\
  \bibnamefont {Poumirol}}, \bibinfo {author} {\bibfnamefont {A.}~\bibnamefont
  {Turchanin}},\ and\ \bibinfo {author} {\bibfnamefont {B.}~\bibnamefont
  {Urbaszek}},\ }\bibfield  {title} {\bibinfo {title} {Exciton spectroscopy and
  unidirectional transport in {MoSe2}-{WSe2} lateral heterostructures
  encapsulated in hexagonal boron nitride},\ }\href
  {https://doi.org/10.1038/s41699-022-00354-0} {\bibfield  {journal} {\bibinfo
  {journal} {npj 2D Materials and Applications}\ }\textbf {\bibinfo {volume}
  {6}},\ \bibinfo {pages} {84} (\bibinfo {year} {2022})}\BibitemShut {NoStop}%
\bibitem [{\citenamefont {Shanks}\ \emph {et~al.}(2022)\citenamefont {Shanks},
  \citenamefont {Mahdikhanysarvejahany}, \citenamefont {Stanfill},
  \citenamefont {Koehler}, \citenamefont {Mandrus}, \citenamefont {Taniguchi},
  \citenamefont {Watanabe}, \citenamefont {LeRoy},\ and\ \citenamefont
  {Schaibley}}]{shanks_interlayer_2022}%
  \BibitemOpen
  \bibfield  {author} {\bibinfo {author} {\bibfnamefont {D.~N.}\ \bibnamefont
  {Shanks}}, \bibinfo {author} {\bibfnamefont {F.}~\bibnamefont
  {Mahdikhanysarvejahany}}, \bibinfo {author} {\bibfnamefont {T.~G.}\
  \bibnamefont {Stanfill}}, \bibinfo {author} {\bibfnamefont {M.~R.}\
  \bibnamefont {Koehler}}, \bibinfo {author} {\bibfnamefont {D.~G.}\
  \bibnamefont {Mandrus}}, \bibinfo {author} {\bibfnamefont {T.}~\bibnamefont
  {Taniguchi}}, \bibinfo {author} {\bibfnamefont {K.}~\bibnamefont {Watanabe}},
  \bibinfo {author} {\bibfnamefont {B.~J.}\ \bibnamefont {LeRoy}},\ and\
  \bibinfo {author} {\bibfnamefont {J.~R.}\ \bibnamefont {Schaibley}},\
  }\bibfield  {title} {\bibinfo {title} {Interlayer {Exciton} {Diode} and
  {Transistor}},\ }\href {https://doi.org/10.1021/acs.nanolett.2c01905}
  {\bibfield  {journal} {\bibinfo  {journal} {Nano Letters}\ }\textbf {\bibinfo
  {volume} {560}},\ \bibinfo {pages} {6599} (\bibinfo {year}
  {2022})}\BibitemShut {NoStop}%
\bibitem [{\citenamefont {Dirnberger}\ \emph {et~al.}(2021)\citenamefont
  {Dirnberger}, \citenamefont {Ziegler}, \citenamefont {Faria~Junior},
  \citenamefont {Bushati}, \citenamefont {Taniguchi}, \citenamefont {Watanabe},
  \citenamefont {Fabian}, \citenamefont {Bougeard}, \citenamefont {Chernikov},\
  and\ \citenamefont {Menon}}]{dirnberger_quasi-1d_2021}%
  \BibitemOpen
  \bibfield  {author} {\bibinfo {author} {\bibfnamefont {F.}~\bibnamefont
  {Dirnberger}}, \bibinfo {author} {\bibfnamefont {J.~D.}\ \bibnamefont
  {Ziegler}}, \bibinfo {author} {\bibfnamefont {P.~E.}\ \bibnamefont
  {Faria~Junior}}, \bibinfo {author} {\bibfnamefont {R.}~\bibnamefont
  {Bushati}}, \bibinfo {author} {\bibfnamefont {T.}~\bibnamefont {Taniguchi}},
  \bibinfo {author} {\bibfnamefont {K.}~\bibnamefont {Watanabe}}, \bibinfo
  {author} {\bibfnamefont {J.}~\bibnamefont {Fabian}}, \bibinfo {author}
  {\bibfnamefont {D.}~\bibnamefont {Bougeard}}, \bibinfo {author}
  {\bibfnamefont {A.}~\bibnamefont {Chernikov}},\ and\ \bibinfo {author}
  {\bibfnamefont {V.~M.}\ \bibnamefont {Menon}},\ }\bibfield  {title} {\bibinfo
  {title} {Quasi-{1D} exciton channels in strain-engineered {2D} materials},\
  }\href {https://doi.org/10.1126/sciadv.abj3066} {\bibfield  {journal}
  {\bibinfo  {journal} {npj 2D Materials and Applications}\ }\textbf {\bibinfo
  {volume} {7}},\ \bibinfo {pages} {eabj3066} (\bibinfo {year}
  {2021})}\BibitemShut {NoStop}%
\bibitem [{\citenamefont {Najafidehaghani}\ \emph {et~al.}(2021)\citenamefont
  {Najafidehaghani}, \citenamefont {Gan}, \citenamefont {George}, \citenamefont
  {Lehnert}, \citenamefont {Ngo}, \citenamefont {Neumann}, \citenamefont
  {Bucher}, \citenamefont {Staude}, \citenamefont {Kaiser}, \citenamefont
  {Vogl}, \citenamefont {Hübner}, \citenamefont {Kaiser}, \citenamefont
  {Eilenberger},\ and\ \citenamefont {Turchanin}}]{najafidehaghani_1d_2021}%
  \BibitemOpen
  \bibfield  {author} {\bibinfo {author} {\bibfnamefont {E.}~\bibnamefont
  {Najafidehaghani}}, \bibinfo {author} {\bibfnamefont {Z.}~\bibnamefont
  {Gan}}, \bibinfo {author} {\bibfnamefont {A.}~\bibnamefont {George}},
  \bibinfo {author} {\bibfnamefont {T.}~\bibnamefont {Lehnert}}, \bibinfo
  {author} {\bibfnamefont {G.~Q.}\ \bibnamefont {Ngo}}, \bibinfo {author}
  {\bibfnamefont {C.}~\bibnamefont {Neumann}}, \bibinfo {author} {\bibfnamefont
  {T.}~\bibnamefont {Bucher}}, \bibinfo {author} {\bibfnamefont
  {I.}~\bibnamefont {Staude}}, \bibinfo {author} {\bibfnamefont
  {D.}~\bibnamefont {Kaiser}}, \bibinfo {author} {\bibfnamefont
  {T.}~\bibnamefont {Vogl}}, \bibinfo {author} {\bibfnamefont {U.}~\bibnamefont
  {Hübner}}, \bibinfo {author} {\bibfnamefont {U.}~\bibnamefont {Kaiser}},
  \bibinfo {author} {\bibfnamefont {F.}~\bibnamefont {Eilenberger}},\ and\
  \bibinfo {author} {\bibfnamefont {A.}~\bibnamefont {Turchanin}},\ }\bibfield
  {title} {\bibinfo {title} {{1D} \textit{p–n} {Junction} {Electronic} and
  {Optoelectronic} {Devices} from {Transition} {Metal} {Dichalcogenide}
  {Lateral} {Heterostructures} {Grown} by {One}‐{Pot} {Chemical} {Vapor}
  {Deposition} {Synthesis}},\ }\href {https://doi.org/10.1002/adfm.202101086}
  {\bibfield  {journal} {\bibinfo  {journal} {Advanced Functional Materials}\
  }\textbf {\bibinfo {volume} {31}},\ \bibinfo {pages} {2101086} (\bibinfo
  {year} {2021})},\ \bibinfo {note} {publisher: American Chemical
  Society}\BibitemShut {NoStop}%
\bibitem [{\citenamefont {Paradisanos}\ \emph {et~al.}(2020)\citenamefont
  {Paradisanos}, \citenamefont {Shree}, \citenamefont {George}, \citenamefont
  {Leisgang}, \citenamefont {Robert}, \citenamefont {Watanabe}, \citenamefont
  {Taniguchi}, \citenamefont {Warburton}, \citenamefont {Turchanin},
  \citenamefont {Marie}, \citenamefont {Gerber},\ and\ \citenamefont
  {Urbaszek}}]{paradisanos_controlling_2020}%
  \BibitemOpen
  \bibfield  {author} {\bibinfo {author} {\bibfnamefont {I.}~\bibnamefont
  {Paradisanos}}, \bibinfo {author} {\bibfnamefont {S.}~\bibnamefont {Shree}},
  \bibinfo {author} {\bibfnamefont {A.}~\bibnamefont {George}}, \bibinfo
  {author} {\bibfnamefont {N.}~\bibnamefont {Leisgang}}, \bibinfo {author}
  {\bibfnamefont {C.}~\bibnamefont {Robert}}, \bibinfo {author} {\bibfnamefont
  {K.}~\bibnamefont {Watanabe}}, \bibinfo {author} {\bibfnamefont
  {T.}~\bibnamefont {Taniguchi}}, \bibinfo {author} {\bibfnamefont {R.~J.}\
  \bibnamefont {Warburton}}, \bibinfo {author} {\bibfnamefont {A.}~\bibnamefont
  {Turchanin}}, \bibinfo {author} {\bibfnamefont {X.}~\bibnamefont {Marie}},
  \bibinfo {author} {\bibfnamefont {I.~C.}\ \bibnamefont {Gerber}},\ and\
  \bibinfo {author} {\bibfnamefont {B.}~\bibnamefont {Urbaszek}},\ }\bibfield
  {title} {\bibinfo {title} {Controlling interlayer excitons in {MoS2} layers
  grown by chemical vapor deposition},\ }\href
  {https://doi.org/10.1038/s41467-020-16023-z} {\bibfield  {journal} {\bibinfo
  {journal} {Nature Communications}\ }\textbf {\bibinfo {volume} {11}},\
  \bibinfo {pages} {2391} (\bibinfo {year} {2020})}\BibitemShut {NoStop}%
\bibitem [{\citenamefont {Poumirol}\ \emph {et~al.}(2020)\citenamefont
  {Poumirol}, \citenamefont {Paradisanos}, \citenamefont {Shree}, \citenamefont
  {Agez}, \citenamefont {Marie}, \citenamefont {Robert}, \citenamefont
  {Mallet}, \citenamefont {Wiecha}, \citenamefont {Larrieu}, \citenamefont
  {Larrey}, \citenamefont {Fournel}, \citenamefont {Watanabe}, \citenamefont
  {Taniguchi}, \citenamefont {Cuche}, \citenamefont {Paillard},\ and\
  \citenamefont {Urbaszek}}]{poumirol_unveiling_2020}%
  \BibitemOpen
  \bibfield  {author} {\bibinfo {author} {\bibfnamefont {J.-M.}\ \bibnamefont
  {Poumirol}}, \bibinfo {author} {\bibfnamefont {I.}~\bibnamefont
  {Paradisanos}}, \bibinfo {author} {\bibfnamefont {S.}~\bibnamefont {Shree}},
  \bibinfo {author} {\bibfnamefont {G.}~\bibnamefont {Agez}}, \bibinfo {author}
  {\bibfnamefont {X.}~\bibnamefont {Marie}}, \bibinfo {author} {\bibfnamefont
  {C.}~\bibnamefont {Robert}}, \bibinfo {author} {\bibfnamefont
  {N.}~\bibnamefont {Mallet}}, \bibinfo {author} {\bibfnamefont {P.~R.}\
  \bibnamefont {Wiecha}}, \bibinfo {author} {\bibfnamefont {G.}~\bibnamefont
  {Larrieu}}, \bibinfo {author} {\bibfnamefont {V.}~\bibnamefont {Larrey}},
  \bibinfo {author} {\bibfnamefont {F.}~\bibnamefont {Fournel}}, \bibinfo
  {author} {\bibfnamefont {K.}~\bibnamefont {Watanabe}}, \bibinfo {author}
  {\bibfnamefont {T.}~\bibnamefont {Taniguchi}}, \bibinfo {author}
  {\bibfnamefont {A.}~\bibnamefont {Cuche}}, \bibinfo {author} {\bibfnamefont
  {V.}~\bibnamefont {Paillard}},\ and\ \bibinfo {author} {\bibfnamefont
  {B.}~\bibnamefont {Urbaszek}},\ }\bibfield  {title} {\bibinfo {title}
  {Unveiling the {Optical} {Emission} {Channels} of {Monolayer}
  {Semiconductors} {Coupled} to {Silicon} {Nanoantennas}},\ }\href
  {https://doi.org/10.1021/acsphotonics.0c01175} {\bibfield  {journal}
  {\bibinfo  {journal} {ACS Photonics}\ }\textbf {\bibinfo {volume} {7}},\
  \bibinfo {pages} {3106} (\bibinfo {year} {2020})}\BibitemShut {NoStop}%
\bibitem [{\citenamefont {Shimasaki}\ \emph {et~al.}(2022)\citenamefont
  {Shimasaki}, \citenamefont {Nishihara}, \citenamefont {Matsuda},
  \citenamefont {Endo}, \citenamefont {Takaguchi}, \citenamefont {Liu},
  \citenamefont {Miyata},\ and\ \citenamefont
  {Miyauchi}}]{shimasaki_directional_2022}%
  \BibitemOpen
  \bibfield  {author} {\bibinfo {author} {\bibfnamefont {M.}~\bibnamefont
  {Shimasaki}}, \bibinfo {author} {\bibfnamefont {T.}~\bibnamefont
  {Nishihara}}, \bibinfo {author} {\bibfnamefont {K.}~\bibnamefont {Matsuda}},
  \bibinfo {author} {\bibfnamefont {T.}~\bibnamefont {Endo}}, \bibinfo {author}
  {\bibfnamefont {Y.}~\bibnamefont {Takaguchi}}, \bibinfo {author}
  {\bibfnamefont {Z.}~\bibnamefont {Liu}}, \bibinfo {author} {\bibfnamefont
  {Y.}~\bibnamefont {Miyata}},\ and\ \bibinfo {author} {\bibfnamefont
  {Y.}~\bibnamefont {Miyauchi}},\ }\bibfield  {title} {\bibinfo {title}
  {Directional {Exciton}-{Energy} {Transport} in a {Lateral} {Heteromonolayer}
  of {WSe} $_{\textrm{2}}$ –{MoSe} $_{\textrm{2}}$},\ }\href
  {https://doi.org/10.1021/acsnano.2c01890} {\bibfield  {journal} {\bibinfo
  {journal} {ACS Nano}\ }\textbf {\bibinfo {volume} {16}},\ \bibinfo {pages}
  {8205} (\bibinfo {year} {2022})}\BibitemShut {NoStop}%
\bibitem [{\citenamefont {Alosious}\ \emph {et~al.}(2020)\citenamefont
  {Alosious}, \citenamefont {Kannam}, \citenamefont {Sathian},\ and\
  \citenamefont {Todd}}]{alosious_kapitza_2020}%
  \BibitemOpen
  \bibfield  {author} {\bibinfo {author} {\bibfnamefont {S.}~\bibnamefont
  {Alosious}}, \bibinfo {author} {\bibfnamefont {S.~K.}\ \bibnamefont
  {Kannam}}, \bibinfo {author} {\bibfnamefont {S.~P.}\ \bibnamefont
  {Sathian}},\ and\ \bibinfo {author} {\bibfnamefont {B.~D.}\ \bibnamefont
  {Todd}},\ }\bibfield  {title} {\bibinfo {title} {Kapitza resistance at
  water–graphene interfaces},\ }\href {https://doi.org/10.1063/5.0009001}
  {\bibfield  {journal} {\bibinfo  {journal} {The Journal of Chemical Physics}\
  }\textbf {\bibinfo {volume} {152}},\ \bibinfo {pages} {224703} (\bibinfo
  {year} {2020})}\BibitemShut {NoStop}%
\bibitem [{\citenamefont {Pollack}(1969)}]{pollack_kapitza_1969}%
  \BibitemOpen
  \bibfield  {author} {\bibinfo {author} {\bibfnamefont {G.~L.}\ \bibnamefont
  {Pollack}},\ }\bibfield  {title} {\bibinfo {title} {Kapitza {Resistance}},\
  }\href {https://doi.org/10.1103/RevModPhys.41.48} {\bibfield  {journal}
  {\bibinfo  {journal} {Reviews of Modern Physics}\ }\textbf {\bibinfo {volume}
  {41}},\ \bibinfo {pages} {48} (\bibinfo {year} {1969})}\BibitemShut {NoStop}%
\bibitem [{\citenamefont {Zipfel}\ \emph {et~al.}(2020)\citenamefont {Zipfel},
  \citenamefont {Kulig}, \citenamefont {Perea-Causín}, \citenamefont {Brem},
  \citenamefont {Ziegler}, \citenamefont {Rosati}, \citenamefont {Taniguchi},
  \citenamefont {Watanabe}, \citenamefont {Glazov}, \citenamefont {Malic},\
  and\ \citenamefont {Chernikov}}]{zipfel_exciton_2020}%
  \BibitemOpen
  \bibfield  {author} {\bibinfo {author} {\bibfnamefont {J.}~\bibnamefont
  {Zipfel}}, \bibinfo {author} {\bibfnamefont {M.}~\bibnamefont {Kulig}},
  \bibinfo {author} {\bibfnamefont {R.}~\bibnamefont {Perea-Causín}}, \bibinfo
  {author} {\bibfnamefont {S.}~\bibnamefont {Brem}}, \bibinfo {author}
  {\bibfnamefont {J.~D.}\ \bibnamefont {Ziegler}}, \bibinfo {author}
  {\bibfnamefont {R.}~\bibnamefont {Rosati}}, \bibinfo {author} {\bibfnamefont
  {T.}~\bibnamefont {Taniguchi}}, \bibinfo {author} {\bibfnamefont
  {K.}~\bibnamefont {Watanabe}}, \bibinfo {author} {\bibfnamefont {M.~M.}\
  \bibnamefont {Glazov}}, \bibinfo {author} {\bibfnamefont {E.}~\bibnamefont
  {Malic}},\ and\ \bibinfo {author} {\bibfnamefont {A.}~\bibnamefont
  {Chernikov}},\ }\bibfield  {title} {\bibinfo {title} {Exciton diffusion in
  monolayer semiconductors with suppressed disorder},\ }\href
  {https://doi.org/10.1103/PhysRevB.101.115430} {\bibfield  {journal} {\bibinfo
   {journal} {Physical Review B}\ }\textbf {\bibinfo {volume} {101}},\ \bibinfo
  {pages} {115430} (\bibinfo {year} {2020})}\BibitemShut {NoStop}%
\bibitem [{\citenamefont {Uddin}\ \emph {et~al.}(2022)\citenamefont {Uddin},
  \citenamefont {Higashitarumizu}, \citenamefont {Kim}, \citenamefont {Yi},
  \citenamefont {Zhang}, \citenamefont {Chrzan},\ and\ \citenamefont
  {Javey}}]{uddin_enhanced_2022}%
  \BibitemOpen
  \bibfield  {author} {\bibinfo {author} {\bibfnamefont {S.~Z.}\ \bibnamefont
  {Uddin}}, \bibinfo {author} {\bibfnamefont {N.}~\bibnamefont
  {Higashitarumizu}}, \bibinfo {author} {\bibfnamefont {H.}~\bibnamefont
  {Kim}}, \bibinfo {author} {\bibfnamefont {J.}~\bibnamefont {Yi}}, \bibinfo
  {author} {\bibfnamefont {X.}~\bibnamefont {Zhang}}, \bibinfo {author}
  {\bibfnamefont {D.}~\bibnamefont {Chrzan}},\ and\ \bibinfo {author}
  {\bibfnamefont {A.}~\bibnamefont {Javey}},\ }\bibfield  {title} {\bibinfo
  {title} {Enhanced {Neutral} {Exciton} {Diffusion} in {Monolayer} {WS}
  $_{\textrm{2}}$ by {Exciton}–{Exciton} {Annihilation}},\ }\href
  {https://doi.org/10.1021/acsnano.2c00956} {\bibfield  {journal} {\bibinfo
  {journal} {ACS Nano}\ }\textbf {\bibinfo {volume} {16}},\ \bibinfo {pages}
  {8005} (\bibinfo {year} {2022})}\BibitemShut {NoStop}%
\end{thebibliography}%
\end{document}